# Emotional Sequential Influence Modeling on False Information

Debashis Naskar[0000−0003−1980−0756], Subhashis Das[0000−0001−9663−9009], and
Sara Rodríguez González[0000−0002−3081−5177]

BISITE, Dept. of Computer Science and Automation, USAL, Salamanca, Spain
{debashis,subhashis,sara}@usal.es

**Abstract.** The extensive dissemination of false information in social networks affects netizen's social lives, morals, and behaviours. When a neighbour expresses strong emotions (e.g., fear, anger, excitement) based on a false statement, these emotions can be transmitted to others, especially through interactions on social media. Therefore, exploring the mechanism that explains how an individual's emotions change under the influence of a neighbour's false statement is a practically important task. In this work, we systematically examining the public's personal, interpersonal, and historical emotional influence based on social context, content, and emotional-based features. The contribution of this paper is to build an emotionally infused model called the Emotional-based User Sequential Influence Model+(E-USIM+) to understand users' temporal emotional propagation patterns and predict future emotions against false information.

## 1 Introduction

The existence of false information in Online Social Networks (OSNs) can trigger strong emotions and arouse netizens' feelings of fear and sadness. Emotional contagion paired with false information intentionally misguides the public by disseminating social media postings. Emotions are indeed regarded as a dominant driver of human behaviour, and understanding the role of emotions in spreading false information is crucial for developing effective strategies to combat their negative impact on social media [4]. False information infused with strong emotional elements such as fear, outrage, or excitement are more likely to be shared and spread rapidly across social networks. In this regard, Zaeem et al. [9] observed the significant association between negative sentiment with fake news and positive sentiment with true news. Emotions have a contagious quality and leading individuals to share content that resonates with their emotional state. Researchers [7] believe that emotions embedded in online rumours are essential determinants of the spreading diffusion dynamics in social media. Recently, a study [2] explored how emotionally framed content triggers emotional response posts by social media users and how corresponding users' response posts affect their sharing behaviour on social media.

In addition to understanding emotional relationships in disseminating false information, capturing internal and external influence plays an important role in social networks. The work in reference [10] proposed influence graph in order to significantly differentiate between intentional and unintentional fake news spreaders. In social networks, existing works have been implemented differently in capturing emotional relationships as well as understanding different influential features on social conversation. For example, significant work [1] has been conducted by capturing personal and social identities within social discourse to better understand opinion behaviours. However, to the best of our knowledge, systematically understanding personal and social influential patterns with dynamic emotion detection over rumour/non-rumour has not yet been explored. On the other hand, it is also essential to capture the user's historical influential pattern in order to understand the dynamic nature of the conversation.

To exhibit this work, we propose an emotionally infused model called *Emotional-based User Sequential Influence Model+*(E-USIM+). The purpose of this work is to identify users' future emotions on false information by systematically examining different influential features along with emotional ones and their temporal propagation patterns. By taking advantage of our proposed model, the user's updated emotion can be predicted by taking into account personal, interpersonal and historical influential patterns on social conversation. The effectiveness of the proposed approach is demonstrated by analysing different topics on publicly available Twitter (currently known as X) datasets.

## 2   Methodology

The ultimate goal of the proposed framework is to detect public emotions in rumour/non-rumour conversations by adopting the BERT and GRU methods. Figure 1 illustrates the overall architecture of the proposed model, which consists of four components, namely *feature-based temporal social network*, *emotion-based user influence network*, *semantic feature extraction* and *emotion classification on false information*. All components are connected to three layers, i.e., the input layer, the fusion layer and the output layer. First, the input layer obtains the user's personal interests, initial emotional propagation structure and set of tweets' words along with structural information. Second, in the fusion layer, the GRU and BERT models are used to obtain the feature vector of the user's temporal internal emotional influential information and semantic feature information represented as $C_{u,i}(t)$ and $F_u(i)$. These two feature vectors are fused to obtain a final feature vector set $H_{u,i}(t) = (C_{u,i}(t) \oplus F_u(i))$, where $\oplus$ indicates the vector concatenating operation. Finally, the output layer is passed through the *Softmax* function to predict the emotion classification as an output class label on false information.

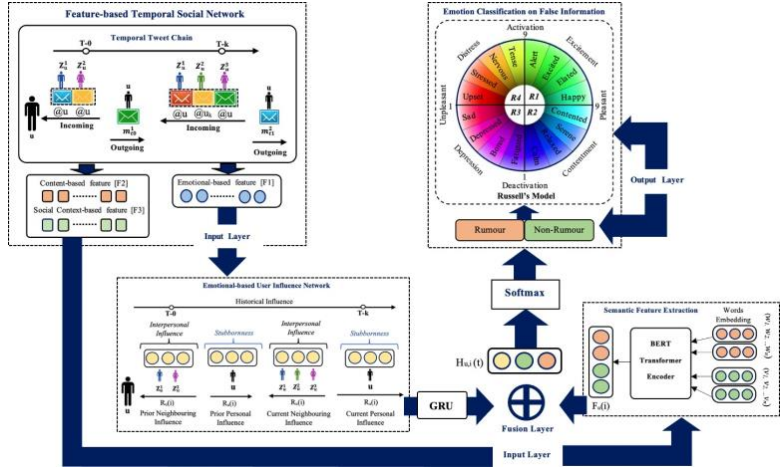

**Fig. 1.** The Overall Architecture of Our Proposed Model

### 2.1 Feature-based Temporal Social Network

To effectively study emotional influence and behavioural impact on social media, we have formulated *feature-based temporal social network* based on temporal conversation by identifying three distinct features, i.e., *emotional-based features [F1]*, *content-based features [F2]* and *social context-based features [F3]*.

Given a direct graph $G = (V, E)$, where each vertex $u \in V$ represents a user and each edge $(u, v) \in E$ represents a user $u$ following to another user $v$. The each user $u \in V$, his/her posting sequence contains $PS_u = < M_u(i), S_u(i), t_u(i) >$ where a tweet message $M_u(i)$ with emotion state $S_u(i)$ at time $t_u(i)$ posted by $u$. Taken into account $G$, we considered a collection of all external set for each user $u \in V$, which denoted as a $Z_u = \{v|(u, v) \in E\}$. The total number of users is $N$ and size of the neighbor set $Z_u$ is $n(u)$. For every participating $u$'s data, we have estimated the *incoming* tweets posted by $v$ that the user $u$ has received between the time of two consecutive *outgoing* tweets is referred to as $t_0$ and $t_1$. Additionally, we also include the incoming tweets posted by $v$ before posting $u$'s first tweet at time $t_0$. Given a user $u$, as we construct a typical temporal tweet chain or tuple chain of $u$'s posted a message at each timestamp $t_u(i)$, where $\downarrow$ denotes (incoming) tweets and $\uparrow$ denotes the (outgoing) tweets.

$$u = < .., \downarrow M^C_{t_{0-1}} >, \uparrow M^C_{t_0}, <\downarrow M^C_{t_{0+1}}, .. >, \uparrow M^C_{t_1}, <\downarrow M^C_{t_{1+1}}, ... >, \uparrow M^C_{t_2}, ...$$

where the user $u$ posts his first tweet at $t_0$ denoted by the tuple $\uparrow M^C_{t_{0-i}}$, $i = 1, 2, ..$ and $C \in \{tweet, reply, retweet\}$. Similarly, incoming tweets between the $Z_u$'s tweets are denoted by the tuple $<\downarrow M^c_{t_{k+0}}, \downarrow M^C_{t_{k+1}}, .. >$.

### 2.2 Emotional-based User Influence Network

The *emotional-based feature [F1]* is determined using Russell's circumflex model of affect [8]. We also have adopted well-grounded emotional lexicons EmoLex [5] to increase the coverage of the emotional words in texts. Each tweet message $M_u(i)$ contains the emotional state $R_i^C$ and the core affect map identified sixteen states ($R_u(i) \in \{1, 2, 3, ..., 16\}$). To reduce the number of emotional states, we consider four quadrants defined in Russell's model and named them as *excitement* (R1), *contentment* (R2), *depression* (R3), and *distress* (R4). All the incoming set of emotion tweets from neighbor are denoted by the tuple $<\downarrow R^C_{t_{k+0}}, \downarrow R^C_{t_{k+1}}, ..>$ and fall between the user's personal emotion tweets $\uparrow R^C_{t_k}$ and $\uparrow R^C_{t_{k+1}}$.

$$u = <.., \downarrow R^C_{i,t_{0-1}}>, \uparrow R^C_{i,t_0}, <\downarrow R^C_{i,t_{0+1}}, ..>\uparrow R^C_{i,t_1}, <\downarrow R^C_{i,t_{1+1}}, ..>, \uparrow R^C_{i,t_2}, ..$$

At each time stamp, four crucial emotion influence factors are consider for shaping one's future dynamics emotion, i.e., prior neighboring $PN_{u,i}(t)$, prior personal $PP_{u,i}(t)$, current neighboring $CN_{u,i}(t)$ and historical emotion influence $HI_{u,i}(t)$. Therefore, future emotion state $S_u(i)$ is classified at the next timestamp $t_u$ based on personal emotion sequence $\uparrow PR_{u,i}(t) = \langle \uparrow R^C_{i,t_0} ..., \uparrow R^C_{i,t_k} \rangle$ sent by the user $u$ at $t$ and all the neighboring emotion of the messages $\downarrow NR_{u,i}(t) = \langle \downarrow R^C_{i,t_{0-1}} ..., \downarrow R^C_{i,t_k} \rangle$ received from $t-1$ to $t$. To proposed the sequential influence model, we already taken place personal and neighboring emotion influence, i.e., derived by $X_{u,i}(t) \in \langle PN_{u,i}(t), PP_{u,i}(t), CN_{u,i}(t) \rangle$. To update $u$'s next emotion at $t_u(i+1)$, the historical emotion influence $HI_{u,i}(t-1)$ could be replaced by the temporal internal emotion state $C_{u,i}(t-1)$. In our sequential influence model, we acknowledge the GRU to prove more appropriate and compact for dropping irrelevant information and updating the relevant information with affordable computation cost. This unit is formally expressed as follows:

$$[r_{u,i}(t) = \sigma(W_r[C_{u,i}(t-1), X_{u,i}(t)] + b_r) \quad (1)$$

$$[g_{u,i}(t) = \sigma(W_g[C_{u,i}(t-1), X_{u,i}(t)] + b_g) \quad (2)$$

$$\tilde{C}_{u,i}(t) = \tanh(W_c[[r_{u,i}(t) \cdot C_{u,i}(t-1), X_{u,i}(t)] + b_c) \quad (3)$$

$$C_{u,i}(t) = [g_{u,i}(t) \cdot \tilde{C}_{u,i}(t) + (1 - [g_{u,i}(t)) \cdot C_{u,i}(t-1) \quad (4)$$

where, $W_r, W_g, W_c \in R^{c \times d_w}$, and $b_r, b_g, b_c \in R^c$ and $d_w$ is the dimension of emotion state region.

### 2.3  Semantic Feature Extraction

We have employed a BERT in our proposed framework to extract semantic features information from *content-based feature [F2]* and *social context-based features [F3]* of the user interactions. The multi-head self-attention mechanism of the BERT model significantly resolves the issue of ignoring meaningful context and extracted feature vectors which accurately represent the literal meaning of the text.

The tweet content within the length $n$ is represented as a set of words $Q=\{w_1, w_2, ..., w_n\}$ and the corresponding social structural feature vector set $K= \{v_1, v_2, ..., v_n\}$ is obtained through the BERT pre-processing model. The initialized text word vectors and structural information are submitted to the Transformer, where feature vector $(Q, K) \in T$ learned by the multi-headed self-attention layer to fed the feed-forward neural network, and the learned tweet semantic representation vector $F_u$ is obtained as follows:

$$F_u(i) = max(0, TW_a + b_a) \qquad (5)$$

where $W_a$ and $b_a$ are the corresponding weight matrix and bias terms of the feed-forward neural network.

### 2.4 Emotion Classification on False Information

For the purpose of significantly detecting $u$'s future emotion state $S_u(i)$ on rumour/non-rumour, we formulate the problem as a predicting probability distribution by given final fusion information $H_{u,i}(t)$ and the output layer for the E-USIM+ is a *softmax* function to output the probabilities of four distinct emotional region over false information.

$$P(S_u(i)|H_{u,i}(t)) = Softmax(V H_{u,i}(t) + b) \qquad (6)$$

where $V \in R^{c \times d_w}$, and $b \in R^c$. $R^c$ is the number of emotional regions for all $u$.

## 3 Experimental Evaluation

We have used a publicly available dataset [3] with 111 events with tweet IDs and user information (60 rumours and 51 non-rumours) to particularly focus on predicting users' future emotions on false information by analyzing user influential features. To understand the dynamic nature of the users, we filtered out from the raw datasets the users who sent 0 or only one message and tweets that did not show any emotion. The final figure after filtering comprises 14,752 users (4876 rumours and 9876 non-rumours) over a total of 1,92,359 tweets posted. For validation of our model, we compare it with other baseline methods, **Degroot** and **Voter** model, along with our simplified version of the **E-USIM** model [6].

Given the sequence of $u$'s emotion, we split the data into 90% of training and the remaining 10% as test dataset according to the temporal sequence. Accuracy (Acc) measures the percentage of correctly predicted emotions among all testing instances. We also compare models in terms of F-measure on *R1-Excitement*, *R2-Contentment*, *R3-Depression*, and *R4-Distress* separately. **Table 1** shows that E-USIM+ almost outperform in all evaluation metrics as compared to other baseline models for both rumour and non-rumour conversations. The result demonstrates that integrating historical information into an influential model is better for predicting someone's emotions than other influential models.

## 4 Conclusion

This paper proposes an emotionally infused model, i.e., E-USIM+, to detect emotion against false information by systematically examining different influential patterns and semantic feature information. Our proposed model has the ability to learn emotional influence and also detect future emotions over false information. The experiments demonstrated that the proposed model outperforms other compared influence models when updating users' future emotions.

**Table 1.** Performance comparison between the E-USIM+ with other Models.

| Method | Rumour | | | | | Non-Rumour | | | | |
|---|---|---|---|---|---|---|---|---|---|---|
| | Acc | F-R1 | F-R2 | F-R3 | F-R4 | Acc | F-R1 | F-R2 | F-R3 | F-R4 |
| DeGroot | 0.5198 | 0.4239 | 0.4811 | 0.5010 | 0.5144 | 0.5802 | 0.6319 | 0.5708 | 0.5213 | 0.4817 |
| Voter | 0.5445 | 0.4286 | 0.4927 | 0.5102 | 0.5024 | 0.5934 | 0.6236 | 0.6199 | 0.5390 | 0.4944 |
| E-USIM | 0.6234 | 0.4982 | 0.5393 | 0.6198 | 0.6305 | 0.6688 | 0.7027 | 0.6910 | 0.6210 | 0.5403 |
| **E-USIM+** | **0.6728** | **0.5699** | **0.5893** | **0.6316** | **0.7012** | **0.7187** | **0.7393** | **0.7245** | **0.6612** | **0.5947** |

**Acknowledgements.** This project has received funding from the EU's Horizon 2020 research and innovation programme under the MSCA grant agreement No. 101034371.